%% file: proton-full-paper.tex
\documentclass[runningheads]{llncs}
\usepackage{bbding}

\usepackage{hyperref}
\usepackage{array}
\usepackage{graphicx}
\usepackage{tabularx}
\usepackage{wrapfig,lipsum,booktabs}
\usepackage{amsmath}
\usepackage{amsfonts}
\usepackage{color}
\usepackage{wrapfig}
\usepackage{algorithmicx}
\usepackage{algorithm}
\usepackage{algpseudocode}
\usepackage{url}
\usepackage{tikz}
\usepackage{amssymb} 
\usepackage{pifont}
\usetikzlibrary{decorations.pathreplacing, positioning, shapes.callouts}
\usepackage{xcolor}
\usepackage{xspace}
\usepackage{float}

\usepackage[most]{tcolorbox}
\tcbuselibrary{listings,skins,many}
\usepackage{color}

\definecolor{dkgreen}{rgb}{0,0.6,0}
\definecolor{gray}{rgb}{0.5,0.5,0.5}
\definecolor{mauve}{rgb}{0.58,0,0.82}

\makeatletter

\def\orcidID#1{\smash{\href{http://orcid.org/#1}{\protect\raisebox{-1.25pt}{\protect\includegraphics{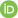}}}}}
\makeatother

\renewcommand\_{\textunderscore\allowbreak}
\newcommand{\bounty}{\textsc{Bounty}\xspace}
\newcommand{\warning}{$\dagger$}

\begin{document}
\title{A shallow dive into the depths of non-termination checking for C programs}
\author{
    Ravindra Metta\inst{1,2}(\Envelope)\orcidID{0000-0001-7368-2389}\and
    Hrishikesh Karmarkar\inst{1}\orcidID{0000-0002-9132-8356}\and
    Kumar Madhukar\inst{3}\orcidID{0000-0001-5686-9758} \and
    R. Venkatesh\inst{1} \and
    Supratik Chakraborty\inst{4}\orcidID{0000-0002-7527-7675}\and
    Samarjit Chakraborty\inst{5}\orcidID{0000-0002-0503-6235}
}
\institute{TCS Research, Tata Consultancy Services, Pune, India \\
    \and
    School of CIT, Technical University of Munich, Munich, Germany \\
    \and
    Dept. of Computer Science, IIT Delhi, New Delhi, India \\
    \and
    Dept. of Computer Science, IIT Bombay, Bombay, India \\
    \and
    Dept. of Computer Science, University of North Carolina, Chapel Hill, USA \\
    \email{r.metta,hrishi.karmarkar,r.venky@tcs.com} \\
    \email{madhukar@cse.iitd.ac.in,supratik@cse.iitb.ac.in,samarjit@cs.unc.edu}
}

\maketitle

\input{abstract}

\section{Introduction}
\label{sec:intr}
\input{intro}

\subsection{Illustrative Example}
\label{sec:example}
\input{example}

\section{Related work}
\label{sec:rela}
\input{relatedwork}


\section{Technique}
\label{sec:technique}

\input{technique}

\subsection{algorithm??}
\label{sec:algo}
\label{sec:algodetails}
\input{algo}

\section{Tool Implementation Details}
\label{sec:tool}
\subsection{Architecture and work flow}
\label{sec:toolimpl}
\input{implementation}
\subsection{Engineering Choices for better performance}
\label{sec:toolimpl}
\input{tooleng}
\subsection{Witness generation}
\label{sec:toolwtiness}
\input{witness-gen}

\section{Experimental Evaluation}
\label{sec:expt}
\input{experiments}

\section{Conclusion and Future Work}
\label{sec:conc}
\input{conclusion}

\newpage

\bibliographystyle{splncs04}
\bibliography{proton-full-paper}

\end{document}

%% file: abstract.tex
\begin{abstract}
Checking for Non-Termination (NT) of a given program $\mathcal{P}$, i.e.,
determining if $\mathcal{P}$ has at least one non-terminating run, is an
undecidable problem that continues to garner significant research attention.
While unintended NT is common in real-world software development,
even the best-performing tools for NT checking are often ineffective on
real-world programs and sometimes incorrect due to unrealistic assumptions such
as absence of overflows. To address this, we propose a sound and efficient
technique for NT checking that is also effective on real-world software.  Given
$\mathcal{P}$, we encode the NT property as an assertion inside each loop of $\mathcal{P}$
to check for recurrent states in that loop, up to a fixed unwinding depth,
using a Bounded Model Checker.  The unwinding depth is increased iteratively
until either NT is found or a predefined limit is reached. Our experiments
on wide ranging software benchmarks show that the technique
outperforms state-of-the-art NT checkers.
\end{abstract}

%% file: intro.tex
The check for Non-Termination (NT) of a given sequential program $\mathcal{P}$
amounts to finding if there exists at least one feasible path in $\mathcal{P}$
that does not terminate.
NT is not only of great theoretical 
interest, but also of significant practical importance. Non-terminating programs
that are undesirably so, particularly in safety- and business-critical software
applications, can lead to serious incidents~\cite{ossincident}.
However, NT checking is well-known to be an undecidable problem
in the general case, meaning that all NT checking techniques are inherently 
\emph{incomplete}. For better scalability, these techniques often impose additional
constraints, such as  restricting the analysis to simple
loops~\cite{10.1007/978-3-030-32304-2_22}, linear
lassos~\cite{10.1007/978-3-319-89963-3_16}, or deterministic
programs~\cite{10.1145/1375581.1375616}. Such restrictions hinder the 
effectiveness of these techniques on a variety of real-world programs.

In a first large-scale empirical study of non-termination bugs in open-source
software, researchers systematically collected 445 non-termination bugs from
over 3,000 GitHub commits~\cite{10.1145/3540250.3549129}. The benchmark
applications included various complex open-source software (OSS) projects, such
as web servers, database management systems, system utilities, networking
tools, and development libraries.  The bugs caused system
crashes, unresponsive interfaces, and server overloads.  The study classified
these bugs based on their root causes, and created a benchmark set with
simplified programs to represent real-world bugs, facilitating a comprehensive
understanding of non-termination issues in real-world software.  
This set includes 56 reproducible non-termination bugs, using which
the authors evaluated state-of-the-art termination analysis tools. The
findings revealed a significant drop in the accuracy of these tools compared
to existing benchmarks, highlighting challenges and limitations in their ability
to handle real-world scenarios.

Recent research~\cite{EndWatch} identified 18 non-termination bugs from
CVE (Common Vulnerability and Exposure) records sourced 
from~\cite{cvewebsite}, encompassing vulnerabilities 
popular multimedia processing, networking, and file
management software. Furthermore, international
competitions like SV-COMP~\cite{SVCOMP24} and Term-Comp~\cite{termcomp2023} feature
dedicated tracks for (non-)termination checking. These competitions
host over 2500 benchmarks, contributed by global academic and
industrial researchers, covering diverse programming features,
 and domains such as reactive systems, product lines, and device drivers.

While leading termination checkers such as UAutomizer~\cite{uauto-svcomp24} and
2LS~\cite{DBLP:conf/tacas/MalikNSV23} perform well on academic
benchmarks~\cite{SVCOMP24}, experimental evaluations highlighted in
\cite{10.1145/3540250.3549129,EndWatch} show a notable decline in their effectiveness and
efficiency when applied to real-world examples.
These evaluations underscore several challenges and limitations faced by the top 
termination analysis tools, including failures in handling complex
 program structures, reliance on assumptions like no-overflows, and overall scalability issues. 
Moreover, existing tools fail to provide
robustness guarantees; if the tools times out, run out of memory, or 
report \emph{unknown}, it is hard to infer the extent of program
state-space explored.
Such insights are crucial for guiding and optimizing subsequent searches,
 so that fruitless explorations maybe avoided.

In our own experience with real-world software, we frequently identify early patterns of program behavior that indicate potential non-termination issues. This observation leads us to hypothesize that exhaustive exploration of program states up to a limited number of loop iterations could reveal these patterns. Bounded Model Checking (BMC)~\cite{DBLP:journals/ac/BiereCCSZ03} is well-established as an efficient technique for such exploration, effectively proving properties of programs up to the bound explored, especially when coupled with SAT and SMT backends~\cite{DBLP:journals/fmsd/ClarkeBRZ01}. However, to our knowledge, BMC has not been applied to non-termination checking. This is due to the fact that non-termination behavior involves infinite traces, whereas BMC examines only bounded-length traces.
 Nevertheless, as argued in this paper, integrating BMC with the concept of
\emph{recurrent states} proposed in~\cite{gupta2008} leads to an efficient and
effective method for checking for non-termination.
A valid program state $S$, which is reachable from some initial state, is recurrent, if
it is feasible to take a path in a program $\mathcal{P}$ along which $S$ repeats.
Then we can take the path fragment from the
repeating state $S$ to itself infinitely many times, thereby proving
the existence of a non-terminating execution.
 In practical terms, in any infinite execution of a real-world program, some state must repeat due to factors such as finite precision in arithmetic operations, which restrict the program's state-space to be finite.
We leverage this principle by inserting
assertions into every loop of $\mathcal{P}$ that assert the absence of
reachable recurrent states. BMC is then employed to verify if any of these
assertions may be violated (i.e., if a recurrent state can be reached). This
approach, which we term {\bounty},  is \emph{complete} modulo the bound for which BMC scales. Therefore,
it either provides proof of NT if such an assertion gets violated, or a 
guarantees NT is absent within the explored bound.

A preliminary version {\bounty} has been implemented in our freely available tool called
PROTON~\cite{DBLP:conf/tacas/MettaKMVC24}. Currently, {\bounty} does not 
support recursive calls, which is a tool restriction rather than 
a limitation of the underlying technique itself.  In this paper, we conducted a
comprehensive experimental evaluation of {\bounty} on all of SV-COMP 2024's 809 
Loop NT benchmarks, all 44 OSS Loop benchmarks, 20 CVE Benchmarks, and 12
complex benchmarks from Term-Comp, that pose difficulties for existing state-of-the-art
NT checkers. {\bounty} not only discovered NT errors missed by other leading
tool, but also identified them significantly faster.
These findings provide robust empirical support for our hypothesis that a substantial number of NT bugs manifest early in program execution. Moreover, they underscore the effective and efficient application of BMC in detecting such bugs via assertions on recurrent states.

The core contributions of this paper are: (i) {\bounty}: an effective and efficient
technique for checking non-termination, (ii) a detailed description of
its underlying algorithm (Sect.~\ref{sec:technique}), (iii) details of our implementation (Sect.~\ref{sec:tool}), and (iv) the results and insights derived from an extensive experimental evaluation (Sect.~\ref{sec:expt}).
 However, before delving into these aspects, the next section presents an illustrative example.



%% file: example.tex
\lstset{frame=tb,
    language=C,
    aboveskip=3mm,
    belowskip=3mm,
    showstringspaces=false,
    columns=flexible,
    basicstyle={\small\ttfamily},
    numbers=none,
    numberstyle=\tiny\color{gray},
    keywordstyle=\color{blue},
    commentstyle=\color{dkgreen},
    stringstyle=\color{mauve},
    breaklines=true,
    breakatwhitespace=true,
    tabsize=3
}

\lstdefinestyle{style2}{
    numbers=none,
    basicstyle={\small},
    language=C,
}

\newtcblisting[number within=section]{nonumlst}[3][]{
    arc=0pt, outer arc=0pt,
    listing only,
    listing style=style2,
    hbox, enhanced,
    title={#3}
}




\begin{figure}[!t]
    \centering
    \begin{minipage}[t]{0.9\columnwidth}
        \begin{lstlisting}[language=C,caption=Example C program with recurrent state instrumentation, label={listingexample}, numbers=left, escapeinside={(*@}{@*)}]
int main(void){
int callVal = nondet_int();
{if (callVal > 0)  { complexFunction(); }
else {
    int i = 0, b*; 
    *b = malloc(sizeof(int));
    *b = callVal;
    {if ((b - (int*)0) > 0){ (*@\fboxrule0.5pt\fboxsep1pt\fcolorbox{black}{blue!10}{\scriptsize myBool pStored0 = myFalse;}@*)
        for (i = 0; i < 10; i++){ 
        (*@\fboxrule0.5pt\fboxsep1pt\fcolorbox{black}{blue!10}{\scriptsize \parbox{0.85\linewidth}{printf("CBMC Instrumentation line 9"); \\
        myBool flag=nondet\_myBool();static int ocallVal;static int oi;\\ if(pStored0) \\  {CPROVER\_assert(!(ocallVal==callVal \&\& oi==i),"recurrent state found");}\\ if(flag) \{ocallVal=callVal;oi=i;pStored0=myTrue;\}}}@*)
            if (i == 2) { 
                i = -1; 
    }}}}}
    return callVal;
}}
\end{lstlisting}
    \end{minipage}
\end{figure}

To demonstrate the effectiveness of {\bounty} in detecting non-terminating behaviors in programs, we present an illustrative example shown in Listing~\ref{listingexample}.
The code shown inside the boxes has been instrumented by {\bounty} and will be explained later.
 In this example, the program can follow two distinct paths based on the value of \texttt{callVal} obtained from \texttt{nondet\_int()}. If \texttt{callVal > 0}, the program calls \texttt{complexFunction()}, which contains a non-terminating execution. Proving such an execution exists is challenging, and most state-of-the-art tools like \texttt{UAutomizer} and \texttt{2LS} fail at it. For instance, \texttt{complexFunction()} might implement an elevator control software, similar to SV-COMP's \texttt{elevator\_spec13\_product21.cil.c} benchmark, which has been difficult for leading NT checkers to solve over the years. Consequently, tools that first analyze the path containing \texttt{complexFunction()} call for non-termination might fail to prove that \texttt{main()} itself has non-terminating executions.

However, there is another source of non-termination in \texttt{main()} due to the \texttt{for} loop. This loop contains a critical condition: when \texttt{i} equals 2, it resets to -1. This reset causes the value of \texttt{i} in the loop to perpetually cycle through the values 0, 1, 2, -1, 0, and so on. The key observation here is that this non-termination manifests itself within five loop iterations, as the sequence of program states cycles through \texttt{i = 0}, \texttt{i = 1}, \texttt{i = 2}, \texttt{i = -1}, and \texttt{i = 0} repeatedly, while other variables remain constant. This sequence of recurrent states implies the existence of a non-terminating execution path.

Thus, one can hypothesize that a non-termination detection strategy, which
\emph{exhaustively} looks for recurrent states that manifest after a few loop iterations
can outperform techniques that are oblivious to this key observation. {\bounty}
by design, exploits this observation to search for recurrent states at shallow
depths of loop executions by deploying a Bounded Model Checking tool. 
First, it instruments the loops in a program with recurrent state assertions as shown in the boxes.
Then, it employs a BMC
tool to validate these assertions at successively larger
loop unwindings starting from a tiny unwind. An assertion failure at a
particular unwinding depth shows the presence of a recurrent state. We
describe the bounded model checker based approach to detect recurrent states and
hence non-termination in Sect.~\ref{sec:technique}.

%% file: relatedwork.tex
\textbf{SV-COMP 2024} As a preliminary version of {\bounty} itself is implemented
in PROTON, the winner in the Termination category, we
compare {\bounty} only with the other best-performing tools, namely
2LS~\cite{2ls-nt-cc}, UAutomizer~\cite{uauto-svcomp24}, and
CPAchecker~\cite{cpachecker}.
The non-termination analysis of 2LS relies on discovering either a singleton recurrence set or an
unchecked arithmetic progression. UAutomizer constructs and explores an
abstract reachability tree using symbolic execution and SMT solving to identify
counterexamples to termination. CPAchecker, based on Configurable Program Analysis (CPA),
checks for non-termination by transforming the liveness check into a
safety check~\cite{SCHUPPAN200679} to look for a state that is revisited.
 While all the tools perform well on SV-COMP benchmarks,
{\bounty} clearly does much better due to its lightweight NT check encoding
and the eagerness to explore shallow behaviors.

\noindent
\textbf{Finding Recurrent Sets} Deriving a recurrent set to detect
non-terminating loops~\cite{gupta2008,giesl2017analyzing,10.1007/978-3-540-79124-9_11,2ls-nt-cc,cpachecker}, with the help of a constraint solver, is popular as
well as
practical.
Recurrent sets can also help reduce the non-termination check to proving
safety, as shown in~\cite{10.1007/978-3-642-54862-8_11}, which also allows them
to tackle non-deterministic programs. Their search, though, is limited to
linear recurrent sets via Farkas’ lemma, unlike {\bounty} and 2LS where
BMC is used. DynamiTe~\cite{dynamite}, iteratively collects
executions and dynamically learns conditions to refine recurrent sets. However,
their technique is particularly focused on nonlinear programs.
Anant~\cite{anant} introduces live abstractions that can be combined with the
concept of closed recurrence sets to soundly disprove termination.
EndWatch~\cite{EndWatch} instruments linear loops with symbolic State Revisit Conditions
and non-linear loops 
with concrete state revisits, which are later checked during program execution. {\bounty} appears to be a promising
addition to this class of techniques.

\noindent
\textbf{Termination and Non-termination}  Proving non-termination and
termination can also go hand in hand, typically by exchanging information,
although the popular practice is to focus on just one at a time.~\cite{alterm}
proposed to do this for non-recursive programs by alternating between refining
an over-approximation and an under-approximation.
~\cite{10.1007/978-3-319-11737-9_18} proposes a new
logical entailment system for temporal constraints and uses Hoare logic
to prove termination and non-termination in a unified
framework.~\cite{10.1007/978-3-662-49674-9_4} synthesize ranking functions and
prove termination and non-termination of imperative programs.~\cite{dynamite}
also infers ranking functions, from concrete transitive closures, to prove
termination along with its NT checking algorithm, and allows the two algorithms
to mutually inform each other. Though {\bounty} currently does not
interact with a termination checker, its iterative approach lends itself quite
naturally to a set-up where it can use and provide useful information to a
termination analysis running alongside.

\noindent
\textbf{Other techniques} ~\cite{ntacceleration} uses loop acceleration
to prove non-termination of integer programs. 
~\cite{geo-nta} puts forth the idea of
a geometric non-termination argument for linear lasso programs. The geometric
non-termination argument is a finite representation of an infinite execution
that has the form of a sum of several geometric series.
~\cite{nt-max-smt} looks for non-termination witnesses in
the form of quasi-invariants, which are sets of configurations that cannot be
left once they are entered.
  The quasi-invariants must also be reachable, which
is checked using a safety prover.  
~\cite{ntreversal}
also uses a safety prover to argue for NT, relying on a syntactic reversal of
the program’s transition system.
 Most of these techniques are orthogonal to the approach of
    {\bounty}, and it may be useful to add some of these as \emph{alternate}
strategies that {\bounty} may explore based on the resource limitations.

%% file: technique.tex
In Listing (\ref{listingexample}), we motivated the need for a technique that can
exhaustively search for a recurrent state that manifests after a few loop iterations. We now present details of the
    {\bounty} technique that uses program instrumentation along with Bounded Model Checking to do this search.
At its core {\bounty} uses a compiler to instrument the program with assertions,
called Recurrent State Assertions (RSIs), that check for the presence of a recurrent state in the program and then employs
a Bounded Model Checker to check these assertions.

\paragraph{Recurrent State Instrumentation (RSI)} The addition of RSI to the program involves annotating it
with additional code. Intuitively, RSI instruments each in loop the program with the ability to
store the program state in any arbitrary iteration and then check if the stored state recurs
in a subsequent iteration. This is encoded in the form of an assertion, which can be checked by a BMC tool.
RSI is done in two parts : the first part, shown on Line 8 of Listing (\ref{listingexample}),
sets a boolean variable \emph{pStored} to false before the loop head, indicating that the program state has not yet been stored in the current invocation of this loop.
The second part, shown on Line 10, the program state is stored by non-deterministically
setting the boolean variable \textbf{flag} to either \emph{false} or \emph{true}.
If the flag is set to \emph{true}, then the state gets stored in the current unwinding, and sets \emph{pStored} to true to indicate that the program state has been stored.
As this storing is done non-deterministically, it accounts for storing the state in any arbitrary iteration of the loop.
This is in contrast to the 2LS technique, in which the encoding involves explicit comparisons between the current iteration's state with the states in  all previous iterations, to detect if a state recurred. This can lead to a blow up in the size of the encoding.
Thus, our encoding is significantly more succinct and allows a program state to be stored  and compared only once.
In EndWatch technique, one program state is stored for each heuristically chosen interval of loop iterations, starting with an interval of 100 iterations, in order to avoid storing too many states.
In contrast,
the SAT or SMT encoding of the instrumented program generated by \bounty ensures that the program state is stored exactly once symbolically, and hence is expected to be more scalable than
in the case where the states across different iterations are stored separately.

A distinction between the current and the previously stored state is maintained in \bounty by making copies of all the program variables with the prefix ``o'' (for ``old'')
in the stored state. For example, \emph{ocallVal} and \emph{oi} respectively represent the values of the variables
\emph{callVal} and \emph{i} during any of the previous iterations if \emph{pStored} is \emph{true}.
Once \emph{pStored} is set to true, i.e., the program state has already been stored in some previous iteration, the stored state is checked against the
current program state for recurrence using an assertion (\emph{CPROVER\_assertion} in Listing (\ref{listingexample})).
If this assertion can be falsified by the bounded model checker, it shows that the state recurred.
This check is repeated with increasing loop unwind bounds as a parameter to the BMC tool until the assertion is violated, or a predefined unwind or time limit is reached.

\paragraph{Bracing} To ensure correct annotation, the program is first passed through a \emph{bracer}, which ensures that the
addition of the annotation does not result in syntactically incorrect code. For
example, the code fragment from Listing (\ref{listingexample}) on Lines 9,11 and 12, but without
braces {\verb|for (i=0;i<10;i++) if (i==2) i=-1;|} is syntactically valid. However, if the
the RSI block on Line 10 is added immediately after the loop head, it
would result in a syntax error if the body of the \verb|for| loop
is not enclosed in braces. The \emph{bracer} tool ensures that the code is correctly
braced before performing the RSI explained above.

\paragraph{Bounded Model Checking} {\bounty} deploys a BMC engine with
backend SAT and SMT solvers.
We assume that a total of $T$ seconds is available to {\bounty} to complete the analysis of each benchmark program, of which
a budget of $T/2$ seconds is allocated to each of the SAT and SMT solvers.
The bounded model checker is invoked iteratively with increasing loop unwind bounds (LUB) until the RSI is violated.
Initially, the SAT solver is employed to attempt falsification of the RSI with the current LUB.
If the SAT solver successfully validates the RSI, then it implies that there are no recurrent states in the program up to the current LUB.
Each successful validation, consumes a part of the total time budget $T/2$ available to the SAT solver.
However, if the SAT solver is able to successfully violate the RSA then {\bounty} returns and reports that the program is non-terminating.
If the SAT solver is unable to falsify the RSA within the time budget, then the SMT solver is invoked to attempt falsification, again
with a time budget of $T/2$ seconds, plus any time remaining from the SAT solver time budget. Following the same strategy as for the SAT solver, the SMT solver is invoked with increasing LUBs
until the RSA is violated or the time budget is exhausted. However, the initial LUB for the SMT solver is set to the LUB at
which the SAT solver last successfully validated the RSA. This is done to ensure that the SMT solver does not repeat the
same work as the SAT solver. Again, if the SMT solver is able to falsify the RSA then {\bounty} returns and reports that the
program is non-terminating.

%% file: algo.tex

%
\subsection{Phase 1: Instrumentation of RSA Check}
\begin{algorithm}
\caption{Bracing and RSI}\label{alg:bracing_rsi}
\begin{algorithmic}[1]
    \Procedure{BraceAndRSI}{$P$}
        \State \textbf{Input}: Compilable program \( P \)
        \State \textbf{Output}: Program \( P_i \) with RSA instrumentation
        \State Instrument P to ensure all control constructs are enclosed in braces
        \For{\textbf{each} loop \( L \) in \( P \)}
            \State Immediately before \( L \), add \textbf{ bool pStored = false; }
            \State At the beginning of the body of \( L \), add the following code;
            \hspace{10em} \If{\textbf{pStored} == \textbf{true}}
                \State assert (stored-state-S != current-state);
            \EndIf
            \State bool \( \text{flag} = \text{nondet\_bool()} \)
            \If{\textbf{flag} == \textbf{true}}
                \State Store current program state in stored-state-S
                \State \textbf{Set} \( \text{pStored} = \text{true} \)
            \EndIf
        \EndFor
        \State \textbf{return} Program \( P_i \) with RSAs
    \EndProcedure
\end{algorithmic}
\end{algorithm}
\subsection{Phase 2: Iterative BMC Procedure}

\begin{algorithm}
\caption{Iterative BMC Procedure}\label{alg:iterative_bmc}
\begin{algorithmic}[1]
    \State \textbf{Input}: \( P_i \), Unwind List \( \text{UWL} \), Timeout T
    \State Initialize loop unwind bound: \( \text{idx} = 0 \)
    \State Initialize \(  BMC = \text{BMC\_SAT} \)
    \State Initialize remaining time budget \( T_r = \frac{T}{2} \)
    \While{\( (\text{idx} < \text{length}(\text{UWL}) ) and ( \text{not TimeOut}) \)}
        \State Invoke BMC-SAT for \( \text{UWL}[\text{idx}] \) and  \( T_r \)
        \If{RSA is falsified}
            \State \textbf{Report NT and STOP}
        \Else
            \State adjust \( T_r \)
            \If{ (\( T_r \) is zero) or (\( \text{SAT solver error} \)) }
                 \State \( BMC = \text{BMC\_SMT}; T_r = \frac{T}{2}+T_r \)
            \EndIf
        \EndIf
    \EndWhile
    \State Return bounded guarantee if assertion not violated up to explored \( LUB \)
\end{algorithmic}
\end{algorithm}

%% file: implementation.tex
A preliminary version of BOUNTY was earlier implemented in the PROTON tool~\cite{DBLP:conf/tacas/MettaKMVC24}
for SV-COMP 2024. That version is now re-implemented in PROTON with substantial improvements by
(1) supporting both SAT and SMT backends, (2)
better time budget allocation to the backend solvers, (3) complete support for most features of C, except recursion,
arrays, and pointers to non-primitive types, and (4) fixing many of the bugs in the tool implementation.
We give an overview of the implementation below. The tool itself is available for download at
\url{https://github.com/kumarmadhukar/term/tree/main/bounty}.

{\bounty} is built using CBMC v5.95.0~\cite{CBMC} with Z3 4.12.2 ~\cite{moura_z3_2008} and
Glucose Syrup~\cite{glucose_2018} as the backend SMT and SAT solvers respectively.
The Bracer and Instrumenter were implemented in C++ using the clang-14 and llvm-14 libraries.
The Instrumenter traverses the Abstract Syntax Tree
(AST) of the input C program to identify variables that form part of the program state in
each loop body and instruments it as part of the RSA (\ref{sec:technique}).
The entire tool flow of {\bounty} is implemented in a bash shell script.

%% file: tooleng.tex
\begin{enumerate}
    \item Ignore, if no RSA check
    \item Disable all user and built in assertions; that is, enable only RSA assertions
    \item Stop on fail
    \item Choice of unwinds
    \item First Glucose (smaller formulae: lesser translation time and solving time), and then Z3 (bigger formulae, lesser translation time, and less memory requirement)
    \item In Expt section, shall we add memory comparson between Glucose and Z3: by running them separately? Also, mention which solver solved hwo may benchmarks? 
    \item Time budgeting: already explained earlier. Can we make it smarter: how about incremental unwinding? Metta to check this
    \item Witness generation
\end{enumerate}

%% file: witness-gen.tex
\begin{enumerate}
    \item Describe what is a sv-comp grapml witness
    \item Describe CPA witness and UA wtiness styles, and we choose CPA stykle?
    \item instrumentation for witness gen --- show motivating example instrumentation extended with witness gen calls
    \item Describe CBMC trace 
    \item Present the witness gen algorithm via CBMC trace
    \item in the experiments section, shall we also run CPA and UA witness validators for all tools (except EndWatch of course, as it is neither available nor produces witness), and see which ones could be validated? This data is not available for  OSS and CVE. So, we can restrict ourselves to that, Of course, we can also present the data on sv-comp benchmarks too. Not a problem. 
    \item In the experiments section, do we present fine-grained categorwise analysis?
    \item In the experiments section, shall we try experiments on divergent cases of Anant benchmarks: and say that, if machine sematics are assumed, they actually terminate?
\end{enumerate}

%% file: experiments.tex
\textbf{Setup:} We have evaluated \bounty on four different benchmark sets:
complete SV-COMP NT benchmarks, 12 complex benchmarks from Term-Comp, OSS
Benchmarks, and CVEs. All our experiments have been performed on a laptop with
Intel i7 3GHz processor, 16GB RAM, Ubuntu 24.04 OS, with Glucose and Z3 as our
respective SAT and SMT backends. For this experimentation, \bounty successively
tries the unwinds \{2, 3, 4, 10, 12, 20, 40, 100, 1000\} until either NT is
found, or an error (including timeout) occurs. We chose these bound values to
systematically evaluate our observation that NT frequently occurs at shallow
depths. We capped the max unwind at 1000 in our experiments; a higher unwind
can certainly be tried if necessary, but BMC is known to struggle for large
unwindings due to the state-space explosion.
\subsection{Tools chosen for comparison:} We compared \bounty's performance and
effectiveness with the following four state-of-the-art NT checkers, as they are
not only top tools, but also represent diverse techniques for checking NT.

\textbf{2LS:}
2LS~\cite{2ls-nt-cc} is a bit-precise C program analyzer that checks for
non-termination by a combination of synthesizing linear lexicographic rankings
using templates, discovering singleton recurrence sets and identifying
recurrence relations for k-induction proofs.

\textbf{CPAChecker (CPA):}
The 2024 version of CPA, CPAChecker 2.3~\cite{cpachecker}, is a significant
upgrade over its earlier versions. It employs a novel strategy selection to
predict, based on carefully picked program features, a suitable sequence of
techniques including k-induction, data-flow analysis, SMT solving, Craig
interpolation, lazy abstraction, and block-abstraction memoization.

\textbf{Ultimate Automizer (UA):}
UA analyzes NT by constructing and exploring an abstract reachability tree
using symbolic execution and SMT solving to identify potential infinite loops.
As SMT-LIB theories support mathematical integers, UA's SMT-LIB translation is
not precise, especially due to the overapproximation of bitwise operators. This
limitation has now been addressed in its 2024 version~\cite{uauto-svcomp24}.

\textbf{EndWatch:}
EndWatch~\cite{EndWatch} detects non-termination by dynamically checking
program execution. It instruments each loop in the program to periodically
store symbolic state revisit conditions for linear loops, and concrete state
for other loops. Using AFL~\cite{AFL} generated tests, EndWatch monitors if any
stored states recur during execution. This method identifies complex
non-termination scenarios influenced by runtime conditions.

The implementations of the above tools employ smart heuristics to guide their
NT search better. Therefore, comparing \bounty with them allows us for a
thorough evaluation. Of the above, 2LS, UA, and CPA perform both termination
and NT checks together.
Therefore, in order to be fair in our experiments, we ran the these tools with
a 15-minutes timeout wit each tool's options set to their SV-COMP defaults.
\bounty and EndWatch check only NT, but not for termination.
Therefore, we ran \bounty with a time budget of only 7 minutes
(less than half the time given to the other checkers).
 Lastly, we could not run EndWatch as it was not
available for download on the
webpage\footnote{\url{https://sites.google.com/view/endwatch/home}}.  However,
the webpage had EndWatch's experimental results with a 15-minute timeout, on a
seemingly faster 3.9GHz AMD Ryzen processor (than our 3 GHz i7
processor)
with 16 GB RAM and Ubuntu 22.04. Further, these results are only on $\sim$13\% of
SV-COMP's NT benchmarks. However, we simply included these results in our
comparisons, even though our processor is slower, because EndWatch's
experiments showed that it could solve many benchmarks the others could not,
and it also serves as a representative dynamic analysis technique.

\subsection{Description of column headers and symbols used in the tables:}
\textbf{\#:} number of benchmarks correctly solved (found to be NT) by the tool\\
\textbf{Time(s):} time, in seconds, the tool to solve the corresponding benchmark\\
\textbf{$\overline{C}$:} \emph{mean} cyclomatic complexity of the category\\
\textbf{C:} cyclomatic complexity of the benchmark\\
\textbf{Res:} result of the tool on the benchmark -- \textbf{$\checkmark$:} NT 
correctly detected; \textbf{\ding{55}:} incorrectly verified as “terminating”; \textbf{U:} reported UNKNOWN (could not verify NT), \textbf{OM:} Out-of-Memory error, \textbf{TO:} Time-Out error, and \textbf{\warning:} failed due to its internal errors such as parsing errors or exceptions.

\input{totals}

\input{svcomp}
\input{termcomp}
\input{oss}

\input{cve}

%% file: totals.tex
\subsection{Benchmarks and Summary of Experimental Evaluation}
\label{sec:totals}
\input{table-totals}
We evaluated BOUNTY on four different benchmark suites.
\begin{enumerate}
	\item \textbf{SV-COMP 2024} -- We took 809 of the 818 non-terminating
benchmarks from SV-COMP 2024, except for 9 recursive programs as \bounty's
current instrumentation does not support recursion.

	\item \textbf{TermComplex} -- We chose all 12 benchmarks from the
SV-COMP\_Mixed\_Categories category of Term-Comp. These are adapted from
SV-COMP benchmarks to be non-terminating, and the most complex benchmarks of
Term-Comp, which other tools are finding difficult to solve. We refer to them
as \textit{Term-Complex} in the rest of this paper.

	\item \textbf{OSS} -- We chose all the 44 non-terminating benchmarks
with loops cited in~\cite{10.1145/3540250.3549129} (available at
\url{https://github.com/FSE2022benchmarks/termination/tree/main/benchmark/loop})
for evaluation on real world NT errors from open source software.

	\item \textbf{CVE} -- In order to assess \bounty on known
vulnerabilities, we tried to choose all 18 from~\cite{EndWatch}. Unfortunately,
8 of these have compilation errors and hence none of the tools could run on
them. So, we chose the remaining 10 that compiled successfully.
\end{enumerate}

In Table~\ref{tab:totals}, we present the overall evaluation of \bounty across
the four benchmark suites as each of these provides a distinct set of
challenges, designed to assess various aspects of program analysis and
verification tools. SV-COMP focuses on software verification, Term-Comp on
termination analysis, OSS on real-world open-source software, and CVEs on known
vulnerabilities.

Table~\ref{tab:totals} presents a macro view of \bounty's performance, compared
to top NT checkers.  More fine grained results per benchmark suite are
presented in the respective evaluation sections that follow. Here, column C
shows the mean cyclomatic complexity of each of the benchmark suites, with the
entry in the last row showing the mean of these means.

Overall, \bounty could solve about 91\% of the total benchmarks, taking an
average time of 24.6 seconds per benchmark. This is 12\% more benchmarks than
the next best tool 2LS, and also $\sim$2x faster time than 2LS, $\sim$4.8x
faster than CPA, and $\sim$2.6x faster than UA. In the last column on EndWatch,
the numbers in brackets denote the total benchmarks for which EndWatch's
experimental data is available. Even with this partial data, it is clear that
\bounty far outperforms EndWatch, with \bounty taking less time for 875
benchmarks than what EndWatch took for just 178 benchmarks. The results
demonstrate that \bounty
consistently outperforms other tools, particularly in the SV-COMP and CVE
benchmarks. While there is room for improvement in time efficiency compared to
tools like 2LS in specific categories, \bounty's overall effectiveness and
efficiency makes it a robust choice for various benchmark suites.

\input{exp-unwind}

%% file: table-totals.tex
\begin{table}[h!]
\centering
\resizebox{\textwidth}{!}{%
\begin{tabular}{|l|r|r|rr|rr|rr|rr|rr|}
\hline
Benchmark Set & \# & $\overline{C}$ & \multicolumn{2}{c|}{BOUNTY} & \multicolumn{2}{c|}{2LS} & \multicolumn{2}{c|}{CPA} & \multicolumn{2}{c|}{UAutomizer} & \multicolumn{2}{c|}{EndWatch} \\
 & & & Res & Time(s) & Res & Time(s) & Res & Time(s) & Res & Time(s) & Res & Time(s) \\
\hline
SV-COMP & 809 & 295 & 744 & 18821.04 & 685 & 36014.06 & 626 & 90771.97 & 545 & 46865.58 & 105 (112) & 6533.46 \\ \hline
Term-Complex & 12 & 141 & 12 & 14.91 & 2 & 8.45 & 8 & 3404.48 & 8 & 2997.86 & 1 (12) & 9917.56 \\ \hline
OSS\_Bench & 44 & 6 & 37 & 2259.55 & 26 & 5879.8 & 21 & 9195.62 & 21 & 5905.04 & 36 (44) & 7279.95 \\ \hline
CVE & 10 & 17 & 8 & 422.84 & 0 & 901.1 & 1 & 306.84 & 3 & 1162.53 & 10 (10) & - \\ \hline
\midrule
Total & 875 & 115 & 801 & 21518.34 & 713 & 42803.41 & 656 & 103678.91 & 577 & 56931.01 & 152 (178) & 23730.97 \\ \hline
\end{tabular}%
}
\caption{Comparison of BOUNTY with 2LS, CPA, UAutomizer, and EndWatch}
\label{tab:totals}
\end{table}

%% file: exp-unwind.tex
\subsubsection{Evaluation of the shallowness of NT across all benchmarks}
\label{sec:unwind}

\input{table3}


Recall that \bounty stores the program state at the beginning of an arbitrary
iteration of each loop, and checks if the state recurs at the beginning of some
subsequent iteration. In particular, \bounty detects a recurrent state in
unwind 2 only if a reachable state at the beginning of iteration 1 repeats
after the iteration, leading to an infinite loop. Similarly, if \bounty detects
a recurrent state in an unwind of $k$, that means it is possible for a state to
recur in at most $k-1$ iterations. This also paves a way for detecting minimal
(shortest) traces for NT, which may help developers debug their programs
faster. It is easy to use \bounty to arrive at such minimal traces, using a
binary search between 0 and $k$ once NT has been found at an unwinding of $k$.

In Table \ref{tab:unwind}, we provide unwind data across all the experiments we
conducted.  Here, column U2 shows number of benchmarks in which NT could be
found in an unwind of 2, U3 shows number of benchmarks in which NT could be
found in an unwind of 3, and so on.  Column UK shows the number of benchmarks
successfully analyzed up to an unwind of 1000, but where no NT could be found.
Column ERR shows the number of benchmarks for which BOUNTY failed either due to
internal errors or timeouts.

The experimental data supports our hypothesis that NT errors can often be found
at relatively shallow depths. In 75\% of SV-COMP NT benchmarks, Recurrent
States (RS) were found at unwinds 2 and 3 (458 and 149, respectively) .  In
Term-Complex, RS were found in all 12 benchmarks at U2 (100\%). In OSS, 28 out
of 44 were found in a max unwind of 4 (63\%), and in CVEs 7 out of 10 were
found at a max unwind of 3.  Overall, RS could be found in about 75\% of the
benchmarks in a max unwind of 3. Less than 3\% (UK total of 24) of the
benchmarks do not have RS until unwind 1000.  This shows across a variety of
benchmarks and a variety of domains, NT checks rarely need to analyse program
unwindings greater than 1000. This makes a compelling case for an exhaustive
search, like \bounty, at shallow depths and a more sophisticated guided search
at greater depths.

Lastly, in Table \ref{tab:unwind}, column ``No RSA'' shows the number of
benchmarks \bounty could not instrument due to unsupported features such as
recursion, arrays, and struct pointers. While a manual analysis revealed that RS
does exists at small unwinding depth for these benchmarks too, the support for
these features remains a part of our immediate future work.

%% file: table3.tex
\begin{table}[h!]
\centering
\begin{tabular}{|l|r|r|r|r|r|r|r|r|r|r|r|r|}
\hline
UNWIND & U2 & U3 & U4 & U10 & U12 & U20 & U40 & U100 & U1000 & UK & ERR & No RSA \\
\midrule
\hline
SV-Comp   & 458  & 149 & 5 & 24 & 1 & 62 & 3 & 2 & 40 & 21 & 15 & 29 \\
\hline
TermComplex & 12 & 0 & - & - & - & - & - & - & - & - & &  - \\
\hline
OSS     & 19 & 6 & 3 & - & 1 & - & 4 & 1 & 3 & 2 & 5 & - \\
\hline
CVE          & 6 & 1 & - & - & - & - & 1 & - & - & 1 & 1 & - \\
\hline
Total=875 & 495 & 156 & 8 & 24 & 2 & 62 & 8 & 3 & 43 & 24 & 21 & 29 \\
\hline
\end{tabular}
\caption{Shallowness of Unwinds}
\label{tab:unwind}
\end{table}

%% file: svcomp.tex
\subsection{Evaluation on SV-COMP 2024 Benchmarks}
\label{sec:svcomp}
\input{table1}

Table \ref{tab:svcomp} shows the experimental results on 809 SV-COMP NT
programs.  818 non-terminating benchmarks were evaluated in SV-COMP 2024.  Of
these, we removed 9 recursive programs as we do not support recursion at the
moment. These are classified by SV-COMP into the categories as shown in Table
\ref{tab:svcomp}, based on the kind of program features they exercise or the
business domain they are intended for, and constitute larger, complex, and
realistic systems. Majority of the categories consist of smaller programs
contributed by researchers, supposedly challenging for different kinds of
verification techniques. Overall, \bounty far outperformed the other tools in
terms of speed as well as the number of benchmarks solved, as indicated by the
last row (Total), owing the shallowness of the recurrent states as pointed out
in Section~\ref{sec:unwind}.

In SV-COMP, categories with $\overline{C} \geq 50$ in Table \ref{tab:svcomp},
described below, consist of complex examples representing real world software.
Category \emph{eca-rers2012} consists of reactive systems with
Event-Condition-Action sequences.  Category \emph{seq-mthreaded} contains
sequentialized code for Physically Asynchronous Logically Synchronous (PALS)
software for distributed systems and also robotic control software.  Category
\emph{product-lines} consists of programs like email-clients, elevator
management, and pumping systems used in mining environments.  Category
\emph{systemc} consists of programs to simulate embedded hardware systems, such
as memory-slave models and networks.  Category \emph{pysco} consists of
programs generated for secure network transmissions. These are augmented with
intermediate conjectures encoded as verification tasks, that do not properly
reflect the behaviour of the
components\footnote{\url{https://github.com/sosy-lab/sv-benchmarks/blob/master/c/psyco/README.txt}}.
Together these 5 categories have 649 benchmarks, out of which bounty solved
644, except for 5 programs in \emph{seq-mthreaded}. All other tools were
ineffective in one more of these categories such as \emph{systemc} and
\emph{seq-mthreaded}. On \emph{eca-rers2012}, \bounty took twice the time of
2LS, as the SAT call in \bounty's implementation went out of time or memory on
several benchmarks, and had to invoke the SMT solver, thus losing time.

Rest of the categories mostly consist of programs with $\overline{C} \ll 50$, crafted by researchers in various papers, based on algorithms like Dijkstra's algorithm for computing square root, and Cohen's algorithm for printing consecutive cubes.  These are meant to motivate why some NT techniques do not work on features like non-linear computations, and seem to be highly contrived.
In some of these categories \bounty fared poorly, e.g.
\emph{termination-crafted}, \emph{termination-crafted}, and \emph{memsafety-ext}. We
manually analyzed some of the failures, and found the reason to invariably be one of the
following.
\begin{enumerate}
\item \textbf{Unsupported features} -- In categories related to memsafety,
\bounty could not instrument the loops as the loops iterate over struct
pointers.
\item \textbf{Very large unwinds} -- In some cases, the first recurrent state
occurs after an enormously large number of iterations, e.g. it occurs after
$2^{32}$ iterations for Cohen's algorithm in \emph{termination-crafted}.
\item \textbf{Potential benchmarking errors} -- Consider ps2-both-nt.c in
\emph{termination-nla}; it does not terminate only if invariants like $(y*y) -
2*x + y = 0$ hold inside the loops there of. But, these invariants, in our
analysis, do not actually hold due to the overflows caused by unconditional
updates inside the loops such as \emph{$y=y+1$}, whereas SV-COMP NT benchmarks
are expected to not contain overflows! In this sense, \bounty may be useful for
identifying benchmarks that have been wrongly classified to be non-terminating.
\end{enumerate}

%% file: table1.tex
\begin{table}[h!]
\centering
\resizebox{\textwidth}{!}{%
\begin{tabular}{|l|r|r|rr|rr|rr|rr|rr|}
\hline
Category & Total & $\overline{C}$ & \multicolumn{2}{c|}{BOUNTY} & \multicolumn{2}{c|}{2LS} & \multicolumn{2}{c|}{CPAChecker} & \multicolumn{2}{c|}{UAutomizer} & \multicolumn{2}{c|}{EndWatch} \\
 & &  & \# & Time(s) & \# & Time(s) & \# & Time(s) & \# & Time(s) & \# & Time(s) \\
\hline
termination-bwb & 14 & 3 & 11 & 438.95 & 11 & 2701.07 & 11 & 915.54 & 8 & 65.64 & 14 (14) & 9.76 \\ \hline
termination-crafted & 16 & 4 &  5 & 841.08 & 9 & 933.45 & 8 & 616.33 & 14 & 99.69 & 10 (11) & 906.53 \\ \hline
termination-crafted-lit & 5 & 5 & 5 & 0.38 & 5 & 0.4 & 4 & 166.47 & 5 & 19.9 & - & - \\ \hline
termination-restricted-15 & 34 & 4 & 34 & 22.84 & 34 & 37.07 & 32 & 1397.02 & 32 & 206.69 & 34 (34) & 20.16 \\ \hline
termination-nla & 19 & 5 & 4 & 1759.41 & 4 & 12494.91 & 5 & 6955.89 & 4 & 1174.4 & 15 (19) & 3740.72 \\ \hline
termination-memory-alloca & 2 & 7 & 2 & 0.57 & 0 & 0.42 & 0 & 2.26 & 2 & 8.79 & 2 (2) & 2.45 \\ \hline
termination-memory-linkedlists & 8 & 8 & 0 & 1.89 & 0 & 0.49 & 0 & 11.67 & 7 & 79.72 & 4 (4) & 19.22 \\ \hline
termination-15 & 3 & 8 & 3 & 1.12 & 0 & 0.36 & 0 & 3.9 & 3 & 23.79 & 6 (6) & 4.59 \\ \hline
bitvector & 7 & 38 & 7 & 1.52 & 7 & 1.35 & 7 & 12.89 & 7 & 30.17 & - & - \\ \hline
openssl-simplified & 1 & 17 & 1 & 0.2 & 1 & 0.1 & 1 & 1.49 & 1 & 5.5 & - & - \\ \hline
locks & 13 & 37 & 13 & 1.69 & 13 & 1.34 & 13 & 1631.03 & 13 & 89.89 & - & - \\ \hline
eca-rers2012 & 200 & 2156 & 200 & 10021.47 & 199 & 5458.22 & 200 & 1956.49 & 60 & 16457.92 & - & - \\ \hline
psyco & 5 & 4725 & 5 & 69.26 & 4 & 43.43 & 4 & 55.53 & 5 & 36.21 & - & - \\ \hline
ldv-regression & 1 & 25 & 1 & 0.09 & 1 & 0.12 & 1 & 1.38 & 1 & 4.06 & - & - \\ \hline
list-ext2-properties & 2 & 27 & 0 & 0.49 & 0 & 0.28 & 0 & 2.57 & 2 & 53.23 & - & - \\ \hline
loops & 9 & 9 & 4 & 15.14 & 6 & 1.51 & 6 & 907.38 & 5 & 62.63 & 9 (9) & 19.3 \\ \hline
loop-acceleration & 2 & 9 & 2 & 0.16 & 2 & 0.23 & 2 & 2.69 & 2 & 8.15 & 2 (2) & 2.17 \\ \hline
loop-crafted & 1 & 9 & 0 & 0.04 & 1 & 0.11 & 0 & 1.32 & 1 & 3.6 & 1 (1) & 0.41 \\ \hline
loop-invariants & 7 & 6 & 1 & 946.91 & 1 & 5400.11 & 6 & 1613.68 & 1 & 61.64 & 7 (7) & 6.09 \\ \hline
loop-invgen & 2 & 12 & 1 & 0.2 & 1 & 0.28 & 1 & 901.5 & 2 & 31.94 & 1 (2) & 900.74 \\ \hline
loop-lit & 7 & 8 & 6 & 63.07 & 6 & 211.41 & 6 & 917.42 & 6 & 65.12 & 0 (1) & 901.32 \\ \hline
product-lines & 334 & 93 & 334 & 106.83 & 333 & 204.01 & 279 & 9884.35 & 325 & 3302.89 & - & - \\ \hline
systemc & 58 & 154 & 58 & 73.26 & 3 & 6.9 & 40 & 17801.2 & 32 & 2236.18 & - & - \\ \hline
seq-mthreaded & 52 & 247 & 47 & 4453.26 & 44 & 8515.44 & 0 & 45002.92 & 0 & 22683.42 & - & - \\ \hline
memsafety & 3 & 32 & 0 & 0.76 & 0 & 0.67 & 0 & 3.95 & 3 & 13.1 & - & - \\ \hline
memsafety-ext & 4 & 20 & 0 & 0.95 & 0 & 0.46 & 0 & 5.1 & 4 & 41.31 & - & - \\
\midrule
Total=809 & 875 & 295 & 744 & 18821.04 & 713 & 42803.41 & 656 & 103678.91 & 577 & 56931.01 & 152 (178) & 23730.97 \\ \hline
\end{tabular}
}
\caption{Summarized Evaluation on SV-COMP Categories, Term-Comp Complex, OSS, and CVE Benchmarks}
\label{tab:svcomp}
\end{table}

%% file: termcomp.tex
\subsection{Evaluation on TermComplex}
\label{sec:termcomp}
\input{table2}

Term-Comp competition semantics assume mathematical integers, which is not
quite realistic. Further, almost all of the Term-Comp benchmarks evaluated by
EndWatch have a cyclomatic complexity of around 5. Due to these two
reasons, we chose the 12 benchmarks under TermComp's
SV-COMP\_Mixed\_Categories, with their cyclomatic complexity ranging from 38 to
271, with a mean CC of 141, as presented in Table~\ref{tab:termcomp}. 

\bounty was able to detect NT for all the 12 benchmarks by checking an unwind
depth of just 2.  The next best tools are CPA and UA, which could solve 8 each,
whereas EndWatch could solve only one. \bounty was also quicker than all these
tools, except for 2LS which took about half the total time of \bounty, but
could solve only 2 benchmarks.


%% file: table2.tex
\begin{table}[h!]
\centering
\resizebox{\textwidth}{!}{%
\begin{tabular}{|l|r|rr|rr|rr|rr|rr|}
\hline
FILE & C & \multicolumn{2}{c|}{BOUNTY} & \multicolumn{2}{c|}{2LS} & \multicolumn{2}{c|}{CPA} & \multicolumn{2}{c|}{UAutomizer} & \multicolumn{2}{c|}{EndWatch} \\
 & & Res & Time(s) & Res & Time(s) & Res & Time(s) & Res & Time(s) & Res & Time(s) \\
\hline
bist\_cellil.c & 62 &  \checkmark & 0.15 & \checkmark & 7.33 & \checkmark & 1.83 & \checkmark & 7.86 & TO & 900 \\ \hline
mem\_slave\_tlm.1il.c & 205  & \checkmark & 6.00 & U & 0.11 & \checkmark & 1.97 & \checkmark & 8.63 & TO & 900 \\ \hline
pc\_sfifo\_1il.c & 38 & \checkmark & 0.17 & U & 0.08 & \checkmark & 1.7 & \checkmark & 10.25 & TO & 900 \\ \hline
test\_locks\_15 & 47 & \checkmark & 0.21 & \checkmark & 0.15 & \checkmark & 236.34 & \checkmark & 16.26 & TO & 900 \\ \hline
token\_ring.01il.c & 59 & \checkmark & 0.19 & U & 0.07 & \checkmark & 1.54 & \checkmark & 5.17 & TO & 900 \\ \hline
token\_ring.05il.c & 127 & \checkmark & 0.54 & U & 0.08 & \checkmark & 5.58 & \checkmark & 18.71 & TO & 900 \\ \hline
token\_ring.10il.c & 212 & \checkmark & 1.36 & U & 0.13 & U & 787.48 & TO & 901.93 & \checkmark & 17.56 \\ \hline
token\_ring.15il.c & 261 & \checkmark & 2.05 & U & 0.14 & U & 782.02 & TO & 900.04 & TO & 900 \\ \hline
transmitter.01il.c & 50 & \checkmark & 0.16 & U & 0.07 & \checkmark & 1.55 & \checkmark & 121.26 & TO & 900 \\ \hline
transmitter.06il.c & 135 & \checkmark & 0.58 & U & 0.08 & \checkmark & 12.92 & \checkmark & 18.93 & TO & 900 \\ \hline
transmitter.11il.c & 220 & \checkmark & 1.42 & U & 0.1 & U & 782.13 & U & 87.95 & TO & 900 \\ \hline
transmitter.16il.c & 271 & \checkmark & 2.08 & U & 0.11 & U & 789.42 & TO & 900.87 & TO & 900 \\ \hline
\midrule
Total=12 & 141 & 12 & 14.91 & 2 & 8.45 & 8 & 3404.48 & 8 & 2997.86 & 1 (12) & 9917.56 \\ \hline
\end{tabular}%
}
\caption{Evaluation of Complex Benchmarks from Term-Comp}
\label{tab:termcomp}
\end{table}

%% file: oss.tex
\subsection{Evaluation on Open Source Software Benchmarks}
\label{sec:oss}
\input{table5}
The OSS benchmark evaluation data, presented in Table~\ref{tab:oss}, highlights
\bounty's ability to handle complex cases efficiently. With a wide range of
cyclomatic complexities (CC) in the OSS benchmarks, \bounty manages to find NT
errors and complete analysis quickly. There are many instances where BOUNTY
identifies errors in less than a second, where other tools either fail to find
the errors or take longer, such as \emph{Signed\_Overflow\_Error\_3\_NT.c}. Further,
CPA and UA also wrongly mark some of these as terminating. This seems to be
mainly due to improper handling of overflows as mentioned earlier in this paper.
In contrast, \bounty not only works as users expect on these real world example,
but also shows robustness by avoiding exceptions and handling timeouts more
gracefully. Overall, \bounty not only solved more benchmarks than the others,
but did so being at least 2x faster than rest of the tools. 

We also analyzed the cases where \bounty failed to identify NT but other tools
succeeded. Invariably, this is due to
loops in which no state recurs before an exceedingly large unwinding.
\bounty, since it employs BMC, cannot scale to such large number of unwindings.
We plan to augment
\bounty with loop acceleration techniques to solve for such loops.

 

%% file: table5.tex
\begin{table}[h!]
\centering
\resizebox{\textwidth}{!}{%
\begin{tabular}{|l|r|rr|rr|rr|rr|rr|}
\hline
FILE & C & \multicolumn{2}{c|}{BOUNTY} & \multicolumn{2}{c|}{2LS} & \multicolumn{2}{c|}{CPA} & \multicolumn{2}{c|}{UAutomizer} & \multicolumn{2}{c|}{EndWatch} \\
 & & Res & Time(s) & Res & Time(s) & Res & Time(s) & Res & Time(s) & Res & Time(s) \\
\hline
Adding\_Subtracting\_Zero\_1\_NT.c & 4 & \checkmark & 0.07 & \checkmark & 0.08 & \checkmark & 2.2 & \checkmark & 5.79 & \checkmark & 0.72 \\ \hline
Adding\_Subtracting\_Zero\_2\_NT.c & 9 & \checkmark & 0.08 & \checkmark & 0.09 & \checkmark & 2.48 & OM & 18.18 & \checkmark & 1.63 \\ \hline
Adding\_Subtracting\_Zero\_3\_NT.c & 12 & \checkmark & 0.19 & \checkmark & 0.12 & U & 873.2 & OM & 17.25 & \checkmark & 1.61 \\ \hline
Adding\_Subtracting\_Zero\_4\_NT.c & 10 & \checkmark & 0.36 & \checkmark & 0.14 & U & 871.14 & OM & 11.84 & TO & 900 \\ \hline
Adding\_Subtracting\_Zero\_5\_NT.c & 6 & \checkmark & 0.28 & U & 0.08 & \warning & 1.35 & \checkmark & 8.48 & \checkmark & 1.51 \\ \hline
Incorrect\_Bit\_Calculation\_1\_NT.c & 2 & \checkmark & 0.07 & \checkmark & 0.1 & \checkmark & 1.29 & \checkmark & 4.76 & \checkmark & 1.51 \\ \hline
Incorrect\_Bit\_Calculation\_2\_NT.c & 2 & \checkmark & 0.07 & \checkmark & 0.11 & \checkmark & 1.23 & \checkmark & 5.04 & \checkmark & 1.62 \\ \hline
Incorrect\_Bit\_Calculation\_3\_NT.c & 2 & \checkmark & 0.07 & \checkmark & 0.11 & TO & 963.68 & \checkmark & 4.79 & \checkmark & 1.52 \\ \hline
Incorrect\_Control\_Statement\_1\_NT.c & 8 & \checkmark & 13.2 & U & 0.28 & \warning & 1.26 & TO & 900.01 & \checkmark & 6.11 \\ \hline
Incorrect\_Control\_Statement\_2\_NT.c & 7 & \checkmark & 0.47 & \checkmark & 10.06 & \warning & 1.21 & \checkmark & 56.61 & \checkmark & 5.38 \\ \hline
Incorrect\_Initialization\_1\_NT.c & 7 & TO & 420.01 & U & 0.24 & \warning & 1.12 & TO & 900.01 & \checkmark & 2.5 \\ \hline
Incorrect\_Initialization\_2\_NT.c & 12 & \checkmark & 0.6 & \checkmark & 0.16 & \checkmark & 1.41 & \checkmark & 8.3 & TO & 900 \\ \hline
Incorrect\_Initialization\_3\_NT.c & 7 & \checkmark & 62.18 & U & 0.26 & \warning & 1.12 & TO & 900.01 & \checkmark & 1.5 \\ \hline
Incorrect\_Initialization\_4\_NT.c & 7 & \checkmark & 0.1 & \checkmark & 0.14 & \checkmark & 1.49 & \checkmark & 6.03 & \checkmark & 1.58 \\ \hline
Incorrect\_Update\_for\_Loop\_Iterator\_1\_NT.c & 7 & \checkmark & 0.08 & \checkmark & 0.13 & \checkmark & 1.31 & \checkmark & 4.83 & \checkmark & 1.52 \\ \hline
Incorrect\_Update\_for\_Loop\_Iterator\_2\_NT.c & 3 & \checkmark & 26.44 & \checkmark & 1.27 & U & 302.38 & U & 19.69 & \checkmark & 0.51 \\ \hline
Incorrect\_Update\_for\_Loop\_Iterator\_3\_NT.c & 3 & \checkmark & 0.08 & \checkmark & 0.1 & \checkmark & 0.98 & \checkmark & 4.6 & \checkmark & 0.53 \\ \hline
Missing\_Corner-case\_Handling\_1\_NT.c & 7 & \checkmark & 0.08 & \checkmark & 0.12 & U & 575.32 & OM & 6.7 & TO & 900 \\ \hline
Missing\_Corner-case\_Handling\_2\_NT.c & 3 & \checkmark & 0.05 & TO & 900 & \checkmark & 299.11 & \checkmark & 5.35 & \checkmark & 4.21 \\ \hline
Missing\_Corner-case\_Handling\_3\_NT.c & 11 & \checkmark & 0.07 & U & 39.99 & U & 861.52 & \ding{55} & 32.89 & \checkmark & 3.17 \\ \hline
Missing\_Corner-case\_Handling\_4\_NT.c & 9 & \checkmark & 0.64 & U & 0.2 & \warning & 1.3 & TO & 900 & \checkmark & 0.4 \\ \hline
Missing\_Initialization\_1\_NT.c & 3 & \checkmark & 0.14 & \checkmark & 0.13 & \checkmark & 1.35 & \checkmark & 5.54 & \checkmark & 0.62 \\ \hline
Missing\_Iterator\_Update\_1\_NT.c & 8 & U & 13.39 & \checkmark & 0.12 & \checkmark & 1.39 & \checkmark & 5.27 & \checkmark & 2.52 \\ \hline
Missing\_Iterator\_Update\_2\_NT.c & 6 & \checkmark & 0.08 & \checkmark & 0.12 & \checkmark & 299.42 & \checkmark & 8.82 & TO & 900 \\ \hline
Missing\_Iterator\_Update\_3\_NT.c & 9 & \checkmark & 0.54 & U & 0.07 & U & 2.04 & U & 22.06 & TO & 900 \\ \hline
Missing\_Iterator\_Update\_4\_NT.c & 7 & \checkmark & 0.06 & U & 281.58 & U & 301.62 & \checkmark & 5.65 & \checkmark & 1.51 \\ \hline
Missing\_Iterator\_Update\_5\_NT.c & 7 & \checkmark & 0.3 & \checkmark & 0.21 & \warning & 0.87 & \checkmark & 7.74 & \checkmark & 1.35 \\ \hline
Misusing\_Variable\_Type\_1\_NT.c & 4 & TO & 420.00 & U & 75.79 & U & 896.08 & TO & 900 & \checkmark & 0.132 \\ \hline
Misusing\_Variable\_Type\_2\_NT.c & 4 & U & 20.96 & TO & 900 & U & 867.6 & U & 11.04 & TO & 900 \\ \hline
Reusing\_Same\_Loop\_Iterator\_1\_NT.c & 5 & \checkmark & 0.14 & \checkmark & 0.1 & \checkmark & 1.28 & \checkmark & 10.76 & \checkmark & 0.51 \\ \hline
Reusing\_Same\_Loop\_Iterator\_2\_NT.c & 4 & \checkmark & 0.18 & \checkmark & 0.11 & \checkmark & 1.29 & \checkmark & 14.69 & \checkmark & 0.612 \\ \hline
Signed\_Overflow\_Error\_1\_NT.c & 2 & \checkmark & 0.36 & \checkmark & 0.12 & \checkmark & 1.8 & \checkmark & 4.78 & \checkmark & 0.314 \\ \hline
Signed\_Overflow\_Error\_2\_NT.c & 2 & \checkmark & 0.4 & \checkmark & 0.17 & \checkmark & 1.8 & \ding{55} & 5.67 & \checkmark & 1.31 \\ \hline
Signed\_Overflow\_Error\_3\_NT.c & 2 & \checkmark & 0.39 & TO & 900 & \ding{55} & 300.45 & \ding{55} & 6.9 & \checkmark & 5.16 \\ \hline
Type\_Conversion\_in\_Assignment\_1\_NT.c & 9 & \checkmark & 0.27 & TO & 900 & \ding{55} & 557.44 & TO & 900.01 & \checkmark & 4.353 \\ \hline
Type\_Conversion\_in\_Comparison\_1\_NT.c & 2 & TO & 420.00 & TO & 900 & \ding{55} & 301.33 & U & 21.3 & \checkmark & 5.203 \\ \hline
Type\_Conversion\_in\_Comparison\_2\_NT.c & 2 & \checkmark & 16.51 & \checkmark & 7.56 & \checkmark & 300.13 & U & 12.55 & \checkmark & 3.353 \\ \hline
Unsigned\_Wraparound\_Error\_1\_NT.c & 5 & \checkmark & 0.13 & \checkmark & 0.09 & \checkmark & 1.31 & U & 10.13 & \checkmark & 5.943 \\ \hline
Unsigned\_Wraparound\_Error\_2\_NT.c & 3 & \checkmark & 0.43 & \checkmark & 0.2 & \checkmark & 4.24 & U & 14.76 & \checkmark & 3.44 \\ \hline
Unsigned\_Wraparound\_Error\_3\_NT.c & 4 & \checkmark & 0.12 & \checkmark & 0.11 & \checkmark & 1.28 & U & 10.04 & TO & 900 \\ \hline
Unsigned\_Wraparound\_Error\_4\_NT.c & 3 & TO & 420.00 & U & 0.24 & \ding{55} & 300.07 & U & 12.25 & \checkmark & 1.23 \\ \hline
Using\_Erroneous\_Condition\_1\_NT.c & 7 & \checkmark & 0.29 & U & 0.09 & \warning & 0.97 & \checkmark & 6.29 & \checkmark & 4.71 \\ \hline
Using\_Erroneous\_Condition\_2\_NT.c & 3 & \checkmark & 0.05 & TO & 900 & \checkmark & 299.11 & \checkmark & 4.99 & \checkmark & 0.15 \\ \hline
Using\_Erroneous\_Condition\_3\_NT.c & 4 & TO & 420.00 & U & 66.42 & \warning & 1.18 & U & 100.96 & TO & 900 \\ \hline
\midrule
Total=44 & 6 & 37 & 2259.55 & 26 & 5879.8 & 21 & 9195.62 & 21 & 5905.04 & 36 (44) & 7279.95 \\ \hline
\end{tabular}
}
\caption{Evaluation on OSS Benchmarks}
\label{tab:oss}
\end{table}

%% file: cve.tex
\subsection{Evaluation on common real-world vulnerabilities and exposures}
\label{sec:cve}

\input{table4}

There are 18 CVE programs evaluated in~\cite{EndWatch}, made available by the
authors at \url{https://github.com/solidConf/CVE_programs}. The exact timings
of EndWatch for these benchmarks are not available and hence we indicate it
with a \textbf{-}.  Of these, except for the 10 shown in \ref{tab:cve}, the
rest had parsing errors and failed to compile with gcc. Therefore, none of the
tools could be run on them. Of the 10 that did compile, \bounty could show NT
in 8, and classified the remaining two as UNKNOWN. As the Total row indicates,
\bounty outperforms 2LS and UA. CPA took lesser time as it failed to complete
its analysis due to internal errors. Also, these tools failed discover NT in
most of the CVEs. Only EndWatch analyzed all 10 to be NT. But we noticed 2 of
these results to be wrong -- they were terminating programs, as we explain
below.

Upon manual analysis of the 10 CVEs, and we found that the two CVEs classified by \bounty as UNKNOWN are
in fact terminating. The benchmark \emph{CVE-2021-45297.c} always terminates
in 32 iterations, due to a typecast from \emph{unsigned int} to \emph{signed
int} in the while loop condition. We confirmed this using CBMC's loop unwinding
assertion check, which clearly showed that the loop will terminate in at most
32 iterations. Another CVE, \emph{libjpeg/CVE-2022-37768.c}, has hard coded
assignments, which cause the only potentially infinite while loop to exit due
to the \emph{exit(1);} call on line 124. This was also confirmed by CBMC's loop
unwinding assertion check. We re-checked both these examples by executing
the two CVEs (they did not take any inputs; so we could readily compile and execute).
%
%
We also manually analyzed the other 8 CVEs, marked by \bounty as
non-terminating and confirmed that they indeed have non-terminating paths. As
stated in Sect.~\ref{sec:svcomp}, this highlights the robustness of \bounty's
analysis, which can be useful in identifying CVEs wrongly classified as
non-terminating.

%% file: table4.tex
\begin{table}[h!]
\centering
\resizebox{\textwidth}{!}{%
\begin{tabular}{|l|r|rr|rr|rr|rr|rr|}
\hline
FILE & C & \multicolumn{2}{c|}{BOUNTY} & \multicolumn{2}{c|}{2LS} & \multicolumn{2}{c|}{CPA} & \multicolumn{2}{c|}{UAutomizer} & \multicolumn{2}{c|}{EndWatch} \\
 & & Res & Time(s) & Res & Time(s) & Res & Time(s) & Res & Time(s) & Res & Time(s) \\
\hline
gpac/CVE-2021-45297.c (TRUE) & 2 & U & 0.53 & TO & 900 & \checkmark & 299.4 & TO & 900 & \checkmark & - \\ \hline
libjpeg/CVE-2022-35166.c & 7 &  \checkmark & 0.30 & U & 0.09 & \warning & 0.7 & OM & 11.34 & \checkmark & - \\ \hline
libjpeg/CVE-2022-37768.c (TRUE) & 22 & TO & 420.00 & U & 0.16 & \warning & 0.73 & \ding{55} & 12.2 & \checkmark & - \\ \hline
nasm/CVE-2021-45257.c & 27 & \checkmark & 0.30 & U & 0.1 & \warning & 1.23 & \ding{55} & 7.19 & \checkmark & - \\ \hline
pdfresurrect/CVE-2021-3508.c & 7 &  \checkmark & 0.35 & U & 0.15 & \warning & 0.71 & OM & 49.03 & \checkmark & - \\ \hline
picoquic/CVE-2020-24944.c & 27 & \checkmark & 0.73 & U & 0.16 & \warning & 0.73 & U & 165.52 & \checkmark & - \\ \hline
wireshark/CVE-2020-26575.c & 21 & \checkmark & 0.10 & U & 0.09 & \warning & 0.7 & \checkmark & 4.04 & \checkmark & - \\ \hline
wireshark/CVE-2021-4185.c & 18 & \checkmark & 0.33 & U & 0.08 & \warning & 1.16 & \checkmark & 5.06 & \checkmark & - \\ \hline
wireshark/CVE-2022-0586.c & 33 & \checkmark & 0.12 & U & 0.19 & \warning & 0.76 & U & 2.6 & \checkmark & - \\ \hline
zziplib/CVE-2020-18442.c & 9 & \checkmark & 0.08 & U & 0.08 & \warning & 0.72 & \checkmark & 5.55 & \checkmark & - \\ \hline
\midrule
Total=10 & 17 & 8 & 422.84 & 0 & 901.1 & 1 & 306.84 & 3 & 1162.53 & 10 (10) & - \\ \hline
\end{tabular}%
}
\caption{Comparison of BOUNTY with 2LS, CPA, UAutomizer, and EndWatch}
\label{tab:cve}
\end{table}

%% file: conclusion.tex
The paper presents \bounty, an efficient technique for checking non-termination
of C programs, and an extensive experimental evaluation that shows its
effectiveness on a wide range of academic as well as real-world benchmarks.
Unlike many other popular NT checking tools, \bounty does not make any
assumption about lack of overflows, presence of a lasso structure,
linear/non-linear computations in the program, etc. \bounty eagerly and
exhaustively explores shallow program behaviors, in tune with its claim that
non-terminating behaviors frequently manifest very early during execution. This
is convincingly validated by our experiments, which clearly show that \bounty
zips through the shallow behaviours of all kinds of benchmarks, including those
with high complexity, successfully detecting non-termination.  In contrast,
higher complexity benchmarks (higher CC) tend to be more challenging for other
NT checkers. To conclude, \bounty is sound, complete up to the unwind depth
that has been explored, and generalizes very well across different software systems.

We are currently working on
substantial tooling updates: especially support for struct pointers.
In future, we intend to augment \bounty with techniques such as 
k-induction and loop acceleration,
and also the ability to find ranking
functions in order to prove termination.